# Chern Insulator Phase in a Lattice of an Organic Dirac Semimetal with Intracellular Potential and Magnetic Modulations

Toshihito Osada*

*Institute for Solid State Physics, University of Tokyo,*

*5-1-5 Kashiwanoha, Kashiwa, Chiba 277-8581, Japan.*

We demonstrate that a Chern insulator could be realized on a real two-dimensional lattice of an organic Dirac semimetal $\alpha$-(BEDT-TTF)$_2$I$_3$ by introducing potential and magnetic modulations in a unit cell. It is a topologically-nontrivial insulator which shows the quantum Hall effect even at zero magnetic field. We assumed a pattern of site potential and staggered plaquette magnetic flux on the lattice so as to imitate the observed stripe charge ordering pattern. When the magnetic modulation is sufficiently large, the system becomes a Chern insulator, where Berry curvatures around two gapped Dirac cones have the same sign on each band, and one chiral edge state connects the conduction and valence bands at each crystal edge. The present model is an organic version of Haldane's model, which discussed the Chern insulator on the honeycomb lattice with second nearest neighbor couplings.

In 1988, Haldane demonstrated that the quantum Hall effect could occur even without the external magnetic field [1]. He considered the model of the two-dimensional (2D) massless Dirac fermion system on the honeycomb lattice like graphene. Additionally, he introduced the second nearest neighbor coupling, the site potential which breaks space inversion symmetry (SIS), and the staggered flux which breaks time reversal symmetry (TRS) but conserves lattice symmetry. He showed that the off-diagonal conductivity is quantized to $\pm e^2/h$ even at zero external magnetic field. It was the pioneering work on the topologically-nontrivial insulting state with TRS breaking, and had a great influence on early studies on topological insulators [2-5]. At the present day, this state is called the Chern insulator or quantum anomalous Hall insulator. The Chern insulator is characterized by a nonzero integer Chern number, which is the topological number [6, 7], and the existence of the chiral edge state [8]. The Chern insulator have been experimentally realized in thin films of a magnetic topological insulator, Cr-doped $(Bi,Sb)_2Te_3$, where only one Dirac cone of the helical surface state is gapped due to TRS breaking [9].

Generally, the magnetic flux in the crystal lattice is represented by the additional phase factors of transfer integrals which surround the flux, reflecting the Aharonov-Bohm effect. In Haldane's model, the condition of zero external magnetic field rules out any additional phase factor in the nearest neighbor couplings in the honeycomb lattice. So, the second nearest neighbor couplings were introduced to represent the magnetic field modulation inside a unit cell. However, this intracellular flux modulation is some unrealistic and seems hardly realized experimentally because it is too fine.

In this paper, we demonstrate that the Chern insulator could be realized in another real crystal lattice without any additional transfer integrals. We employ a crystal lattice of an organic layered conductor $\alpha$-(BEDT-TTF)$_2$I$_3$, where BEDT-TTF denotes bis(ethylenedithio)-tetrathiafulvalene [10]. It can be regarded as a 2D massless Dirac fermion system, which has a pair of Dirac cones like graphene [11-13]. Moreover, this compound undergoes a first-order phase transition into an insulating state due to charge ordering (CO) at the transition temperature $T_{CO}$, which is about 135 K at ambient pressure and goes down to zero at the critical pressure $P_c \sim 1.1$ GPa [10]. In below, we introduce the realistic site potential and staggered magnetic flux into the 2D lattice of $\alpha$-(BEDT-TTF)$_2$I$_3$, and investigate possible topological phases.

Figure 1(a) shows the schematic of lattice structure of a 2D conducting layer of $\alpha$-(BEDT-TTF)$_2$I$_3$. There are four BEDT-TTF molecules dubbed A, A', B, and C in a unit cell indicated by a dashed rectangle. Their highest occupied molecular orbitals (HOMOs) form four $\pi$-bands. The seven types transfer integrals between HOMOs, $a_1$, $a_2$, $a_3$, $b_1$, $b_2$, $b_3$, and $b_4$, are also indicated in the figure. In the CO insulator phase, a stripe pattern of charge density appears on the lattice [14]. In addition, the same stripe pattern of spin density was also observed just below $T_{CO}$ [15]. As indicated by smeared circles, the charge density and spin density are large at sites A and B, but small at sites A' and C in the stripe pattern.

In our model, we introduce the site potential and staggered flux so as to simulate the observed stripe CO pattern without any additional transfer integrals. As shown in Fig. 1(b), breaking SIS, we allot potentials $\Delta$ and $-\Delta$ to sites A and A', respectively. There are eight triangular plaquettes defined by transfer lines in a unit cell. We allot magnetic fluxes $\Phi$ and $-\Phi$ to plaquettes $\triangle$BAA' and $\triangle$CAA', respectively, which cause difference of magnetic environment between B and C sites. Note that the total flux in a unit cell remains zero. This intracellular flux modulation breaks TRS but conserves lattice symmetry. The plaquette fluxes are represented by adding phase factors $e^{\pm i\theta}$ to transfer integrals $b_1$, $b_2$, $b_3$, and $b_4$. Here, the Peierls phase $\theta$ relates to the plaquette flux $\Phi$ by

$$\theta = \frac{e}{\hbar}\Phi = 2\pi\frac{\Phi}{h/e}.$$

(1)

We consider the conventional tight-binding model for $\alpha$-(BEDT-TTF)$_2$I$_3$ lattice [11, 12].

$$H(\mathbf{k}) = \begin{pmatrix} \varepsilon_0 + \Delta & H_{AA'}(\mathbf{k}) & H_{AB}(\mathbf{k}) & H_{AC}(\mathbf{k}) \\ H_{AA'}(\mathbf{k})^* & \varepsilon_0 - \Delta & H_{A'B}(\mathbf{k}) & H_{A'C}(\mathbf{k}) \\ H_{AB}(\mathbf{k})^* & H_{A'B}(\mathbf{k})^* & \varepsilon_0 & H_{BC}(\mathbf{k}) \\ H_{AC}(\mathbf{k})^* & H_{A'C}(\mathbf{k})^* & H_{BC}(\mathbf{k})^* & \varepsilon_0 \end{pmatrix},$$

(2)

$$H_{AA'}(\mathbf{k}) = a_2 e^{+i\mathbf{k}\cdot\boldsymbol{\tau}_1} + a_3 e^{-i\mathbf{k}\cdot\boldsymbol{\tau}_1},$$

$$H_{AB}(\mathbf{k}) = (b_2 e^{-i\theta})e^{+i\mathbf{k}\cdot\boldsymbol{\tau}_2} + (b_3 e^{+i\theta})e^{-i\mathbf{k}\cdot\boldsymbol{\tau}_3},$$

$$H_{AC}(\mathbf{k}) = (b_1 e^{-i\theta})e^{+i\mathbf{k}\cdot\boldsymbol{\tau}_3} + (b_4 e^{+i\theta})e^{-i\mathbf{k}\cdot\boldsymbol{\tau}_2},$$

$$H_{A'B}(\mathbf{k}) = (b_2 e^{-i\theta})e^{-i\mathbf{k}\cdot\boldsymbol{\tau}_2} + (b_3 e^{+i\theta})e^{+i\mathbf{k}\cdot\boldsymbol{\tau}_3},$$

$$H_{A'C}(\mathbf{k}) = (b_1 e^{-i\theta})e^{-i\mathbf{k}\cdot\boldsymbol{\tau}_3} + (b_4 e^{+i\theta})e^{+i\mathbf{k}\cdot\boldsymbol{\tau}_2},$$

$$H_{BC}(\mathbf{k}) = a_1 e^{+i\mathbf{k}\cdot\boldsymbol{\tau}_1} + a_1 e^{-i\mathbf{k}\cdot\boldsymbol{\tau}_1}.$$

The bases of 4×4 Hamiltonian matrix $H(\mathbf{k})$ are four Bloch sums constructed from HOMOs of A, A', B, and C molecules. $\mathbf{k} = (k_x, k_y)$ is a 2D wave vector, and $\boldsymbol{\tau}_1 = (0, a/2)$, $\boldsymbol{\tau}_2 = (b/2, -a/4)$, and $\boldsymbol{\tau}_3 = (b/2, a/4)$, where $b$ and $a$ are lattice constants along the $x$ and $y$ directions, respectively. The energy of the HOMO, $\varepsilon_0$, is taken as zero. We have chosen the transfer integrals as $a_1 = -0.038$ eV, $a_2 = +0.080$ eV, $a_3 = -0.018$ eV, $b_1 = +0.123$ eV, $b_2 = +0.146$ eV, $b_3 = -0.070$ eV, and $b_4 = -0.025$ eV, so as to obtain the Dirac semimetal dispersion in the case of $\Delta = 0$ and $\theta = 0$ [16]. They correspond to the case of uniaxial pressure $P_a = 0.4$ MPa in previous calculations. The site potentials $\Delta$ and $-\Delta$ are added in diagonal elements of sites A and A', respectively. The phase factors $e^{\pm i\theta}$ are also introduced to transfer integrals $b_1$, $b_2$, $b_3$, and $b_4$.

We can obtain the band dispersion by diagonalization of $H(\mathbf{k})$. The examples of the dispersion of the third (valence) and fourth (conduction) bands are shown in upper panels in Fig. 2 for three parameter sets. The system is a 2D zerogap Dirac semimetal with two Dirac cones at $\Delta = 0$ and $\Phi = 0$. When the site potential $\Delta$ becomes finite, the energy gaps open at two Dirac points as seen in Fig. 2(a), and the system becomes an insulator. On the other hand, when the staggered magnetic flux $\Phi$ becomes finite, the energy gaps also open at two Dirac points as seen in Fig. 2(c). These insulating states have almost the same band structure.

To see the difference of these insulating states, we investigate the Berry curvature of the valence and conduction bands for each case. The $z$-component of Berry curvature can be calculated by the following formula at each $\mathbf{k}$-point.

$$[\mathbf{B}_n(\mathbf{k})]_z = i \sum_{m(\neq n)} \frac{\langle u_{n\mathbf{k}}|\partial H(\mathbf{k})/\partial k_x|u_{m\mathbf{k}}\rangle \langle u_{m\mathbf{k}}|\partial H(\mathbf{k})/\partial k_y|u_{n\mathbf{k}}\rangle - c.c.}{\{E_m(\mathbf{k}) - E_n(\mathbf{k})\}^2}.$$

(3)

Here, $\mathbf{B}_n(\mathbf{k})$, $E_n(\mathbf{k})$, and $|u_{n\mathbf{k}}\rangle$ are the Berry curvature, eigen energy, and four-component eigen vector of the $n$-th band ($n = 1, 2, 3, 4$), respectively. $\partial H(\mathbf{k})/\partial k_x$ and $\partial H(\mathbf{k})/\partial k_y$ are 4×4 matrices obtained by differentiating $H(\mathbf{k})$ formally. "$c.c.$" denotes the complex conjugate of the first term on the numerator.

Generally, the off-diagonal conductivity is not necessarily zero in the band insulator where the Fermi level is located in a gap. Under the electric field **E**, each electron obtains an anomalous velocity $(e/\hbar)\mathbf{E} \times \mathbf{B}_n(\mathbf{k})$ [17]. Since the summation of the anomalous velocity of all occupied states ($n$, **k**) contributes to the off-diagonal transport, the off-diagonal conductivity of 2D insulators is written as the following at zero temperature and zero magnetic field.

$$\sigma_{xy} = -N_{\text{Ch}} \frac{e^2}{h},$$

(4)

where $N_{\text{Ch}}$ is defined as a summation of Berry curvature $\mathbf{B}_n(\mathbf{k})$ of all occupied states.

$$N_{\text{Ch}} = \sum_{\substack{n \\ E_n < E_F}} \frac{1}{2\pi} \iint_{\text{BZ}} [\mathbf{B}_n(\mathbf{k})]_z dk_x dk_y.$$

(5)

It is proven that $N_{\text{Ch}}$, which is called the Chern number, is always takes an integer value including zero [7]. Therefore, the off-diagonal conductivity of 2D insulators (4) is quantized at zero temperature and zero magnetic field [6, 7].

The calculated Berry curvatures are shown in the lower panels in Fig. 2. In the case of finite $\Delta$ shown in Fig. 2(a), the Berry curvature becomes finite and peak-shaped around two valleys, which are formed by gap-opening at the Dirac points. The signs of the two peaks are opposite on each band, leading $N_{\text{Ch}} = 0$. Under the electric field, electron around two valleys obtain the opposite directions of anomalous velocity. The off-diagonal conduction is absent because of the cancellation of two valleys. Alternatively, the valley current, which is the deference of currents of electrons around two valleys, becomes finite, namely, causes the valley Hall effect [18, 19]. The insulator in Fig. 2(a) is a topologically-trivial ordinary insulator.

In contrast, in the case of finite $\Phi$ shown in Fig. 2(c), the Berry curvature also shows peak structures around two valleys, but they have the same sign. Under the in-plane electric field, electrons around two valleys have the anomalous velocity with the same direction, showing the anomalous Hall effect at zero magnetic field. Since the Chern number $N_{Ch}$ becomes equal to $-1$, the off-diagonal conductivity is quantized to $\sigma_{xy} = +e^2/h$ at zero magnetic field. Because of this quantum anomalous Hall effect, we can conclude that the system is not but a topologically-nontrivial Chern insulator.

The ordinary insulator and the Chern insulator are clearly distinguished topological phases by the topological number $N_{Ch}$. Then, what happens at the boundary of these two phases? Figure 2(b) shows the band dispersion and Berry curvature for a parameter set corresponding to the boundary. In the upper panel, we can see that gap closes at one of two valleys forming one Dirac cone. In other words, the system is a 2D zerogap Dirac semimetal with a single Dirac cone. The Berry curvature has a single peak around the gapped valley, and the single Dirac point becomes a singular point. When the one insulator change to another, the system closes a gap and inverts the sign of Berry curvature at one of two valleys, causing a discontinuous jump of $N_{Ch}$.

Next, we investigate the edge state of the finite system. We consider a monolayer nanoribbon of α-(BEDT-TTF)$_2$I$_3$ parallel to the crystalline **a**-axis (*y*-axis), which has two types of edges, AA'-edge and BC-edge, as shown in Fig. 3(d). The energy spectrum of the α-(BEDT-TTF)$_2$I$_3$ nanoribbon was already studied in the Dirac semimetal state [20] and the CO insulating state [21, 22]. We calculate the $k_y$-dispersion of energy spectrum of the nanoribbon with potential and magnetic modulations following the method mentioned in Ref. [20]. The spectra for the same parameters in Fig. 2 are shown in Fig. 3(a), (b), and (c). Isolated branches appearing in the bulk gap, which are indicated by labels "AA' " and "BC", are the edge states localized around the AA'-edge and BC-edge, respectively. In the ordinary insulator phase, each branch connects to one of two bands. It connects to the conduction band in the case of $\Delta > 0$ as seen in Fig. 3(a). In contrast, in the Chern insulator phase, each edge state branch connects the conduction and valence bands, as seen in Fig. 3(c). Each branch crosses the Fermi level at a single point, and carries electric current in one direction so as to surround the nanoribbon. The existence of this chiral edge state is a characteristic feature of the Chern insulator. In the Chern insulator with $N_{Ch} = \pm 1$, one edge state per one edge is expected from the bulk-edge correspondence [23]. In the Dirac semimetal state at the boundary of the ordinary and Chern insulators, each edge state branch connects the single Dirac point and one band as seen in Fig. 3(b), reconnecting edge state branches.

The phase diagram of 2D $\alpha$-(BEDT-TTF)$_2$I$_3$ lattice with potential and magnetic modulations is shown in Fig. 4. The $\Phi$ - $\Delta$ plane is cut into four regions by two lines crossing at the origin. Each line corresponds to the zerogap Dirac semimetal state with a single Dirac cone. The hatched regions correspond to the topologically-nontrivial Chern insulator phases with $N_{Ch} = \pm 1$, which show quantum anomalous Hall effect. The white regions correspond to the topologically-trivial ordinary insulator phase with $N_{Ch} = 0$, which shows the valley Hall effect. The condition for the Chern insulator is given by the following in the parameter region of Fig. 4.

$$\frac{|\Phi|}{h/e} > \frac{|\Delta|}{0.777 \text{ [eV]}}.$$

(6)

Note that the phase diagram is periodic in much wider region of $\Phi$. The diagram shown in Fig. 4 corresponds to the vicinity of the origin in Haldane's phase diagram (Fig. 2 of Ref. [1]).

Finally, we discuss the possibility that a Chern insulator appears in the real $\alpha$-(BEDT-TTF)$_2$I$_3$ system with CO. According to the mean field theory by Kobayashi *et al.*, the CO transition in $\alpha$-(BEDT-TTF)$_2$I$_3$ could be accompanied by breaking of both SIS and TRS, depending on the Coulomb interaction parameters [24]. The breaking of TRS could lift the spin degeneracy of the electronic state resulting in the spin density modulation in a unit cell. In fact, in addition to the stripe charge density modulation [14], the intracellular spin density modulation with the same pattern was observed in the weak CO state just below $T_{CO}$ by $^{13}$C-NMR measurement [15]. At low temperatures, the spin density modulation disappears due to the formation of spin singlet pairs. So, we focus on the weak CO state. The potential and flux pattern of the present model meets the observed charge and spin density pattern of the weak CO state: The charge density is different between sites A and A', and the magnetic environment is different between sites B and C.

However, it seems difficult for the weak CO state in $\alpha$-(BEDT-TTF)$_2$I$_3$ to satisfy the Chern insulator condition (6). Generally, a magnetic flux penetrating a circle with a radius $r$ surrounding a vertical magnetic dipole **m** is given by $\Phi = \mu_0|\mathbf{m}|/(2r)$ Since the effective magnetic dipole localized at a site never exceeds the Bohr magneton $\mu_B$ in the weak CO state, the upper limit of $|\Phi|/(h/e)$ is estimated as the order of $10^{-5}$ by setting $|\mathbf{m}| < \mu_B$ and $r \sim a/4 = 0.25$ nm. So, the potential modulation amplitude $|\Delta|$ must be much smaller than 0.01 meV to satisfy (6). This value seems rather small in the weak

CO state in real $\alpha$-(BEDT-TTF)$_2$I$_3$ system. The weak CO state just below the critical pressure $P_c \sim 1.1$ GPa might be hopeful. The anomalous metallic transport observed in this region [25] could be explained by the chiral edge transport in the Chern insulator state. It might be also possible that the fluctuation of charge and spin density causes some signature of a Chern insulator around the CO transition.

In summary, we have constructed an organic version of Haldane's model, which first discussed the Chern insulator in the honeycomb lattice. We considered a real 2D lattice of an organic Dirac fermion system $\alpha$-(BEDT-TTF)$_2$I$_3$ without any unrealistic additional transfers, and introduced a pattern of intracellular potential and magnetic modulations imitating its weak CO state. By investigating the Berry curvature and edge states of the system, we show that the Chern insulator appears when the magnetic modulation is relatively large compared to the charge modulation. The realization of the Chern insulator seems not so easy, but might be possible in the weak CO state close to the CO phase boundary in $\alpha$-(BEDT-TTF)$_2$I$_3$. The present model gives an example that a Chern insulator can be considered in a real bulk material.

**Acknowledgements**

The author thanks Mr. K. Yoshimura, Dr. M. Sato, and Dr. T. Taen for valuable discussions. He also thanks Dr. K. Miyagawa, Prof. A. Kobayashi, and Prof. H. Fukuyama for their useful comments and suggestions. This work was supported by JSPS KAKENHI Grant Numbers JP25107003 and JP16H03999.

*corresponding author, osada@issp.u-tokyo.ac.jp

[1] F. D. M. Haldane, Phys. Rev. Lett. **61**, 2015 (1988).

[2] C. L. Kane and E. J. Mele, Phys. Rev. Lett. **95**, 226801 (2005).

[3] L. Fu and C. L. Kane, Phys. Rev. B **74**, 195312 (2006).

[4] X.-L. Qi, Y.-S. Wu, and S.-C. Zhang, Phys. Rev. B **74**, 085308 (2006).

[5] X.-L. Qi, T. L. Hughes, and S.-C. Zhang, Phys. Rev. B **78**, 195424 (2008).

[6] D. J. Thouless, M. Kohmoto, M. P. Nightingale, and M. den Nijs, Phys. Rev. Lett. **49**, 405 (1982).

[7] M. Kohmoto, Ann. Phys. **160**, 343 (1985).

[8] N. Hao, P. Zhang, Z. Wang, W. Zhang, and Y. Wang, Phys. Rev. B **78**, 075438 (2008).

[9] Cui-Zu Chang, *et al.*, Science **340**, 167 (2013).

[10] N. Tajima, S. Sugawara, M. Tamura, Y. Nishio, and K. Kajita, J. Phys. Soc. Jpn. **75**, 051010 (2006).

[11] K. Kajita, Y. Nishio, N. Tajima, Y. Suzumura, and A. Kobayashi, J. Phys. Soc. Jpn. **83**, 072002 (2014).

[12] S. Katayama, A. Kobayashi, and Y. Suzumura, J. Phys. Soc. Jpn. **75**, 054705 (2006).

[13] A. Kobayashi, S. Katayama, Y. Suzumura, and H. Fukuyama, J. Phys. Soc. Jpn. **76**, 034711 (2007).

[14] T. Kakiuchi, Y. Wakabayashi, H, Sawa, T. Takahashi, and T. Nakamura, J. Phys. Soc. Jpn. **76**, 113702 (2007).

[15] K. Ishikawa, M. Hirata, D. Liu, K. Miyagawa, M. Tamura, and K. Kanoda, Phys. Rev. B **94**, 085154 (2016).

[16] R. Kondo, S. Kagoshima, and J. Harada, Rev. Sci. Instrum. **76**, 093902 (2005).

[17] G. Sundaram and Q. Niu, Phys. Rev. **B59**, 14915 (1999).

[18] D. Xiao, W. Yao, and Q. Niu, Phys. Rev. Lett. **99**, 236809 (2007).

[19] G. Matsuno, Y. Omori, T. Eguchi, and A. Kobayashi, J. Phys. Soc. Jpn. **85**, 094710 (2016).

[20] Y. Hasegawa and K. Kishigi, J. Phys. Soc. Jpn. **80**, 054707 (2011).

[21] Y. Omori, G. Matsuno, and A. Kobayashi, JPS Conf. Proc. **1**, 012119 (2014).

[22] Y. Omori, G. Matsuno, and A. Kobayashi, J. Phys. Soc. Jpn. **86**, 074708 (2017).

[23] Y. Hatsugai, Phys. Rev. Lett. **71**, 3697 (1993).

[24] A. Kobayashi, Y. Suzumura, F. Piéchon, and G. Montambaux, Phys. Rev. B **84**, 075450 (2011).

[25] D. Liu, K. Ishikawa, R. Takehara, K. Miyagawa, M. Tamura, and K. Kanoda, Phys. Rev. Lett. **116**, 226401 (2016).

**Figure 1** (Osada)

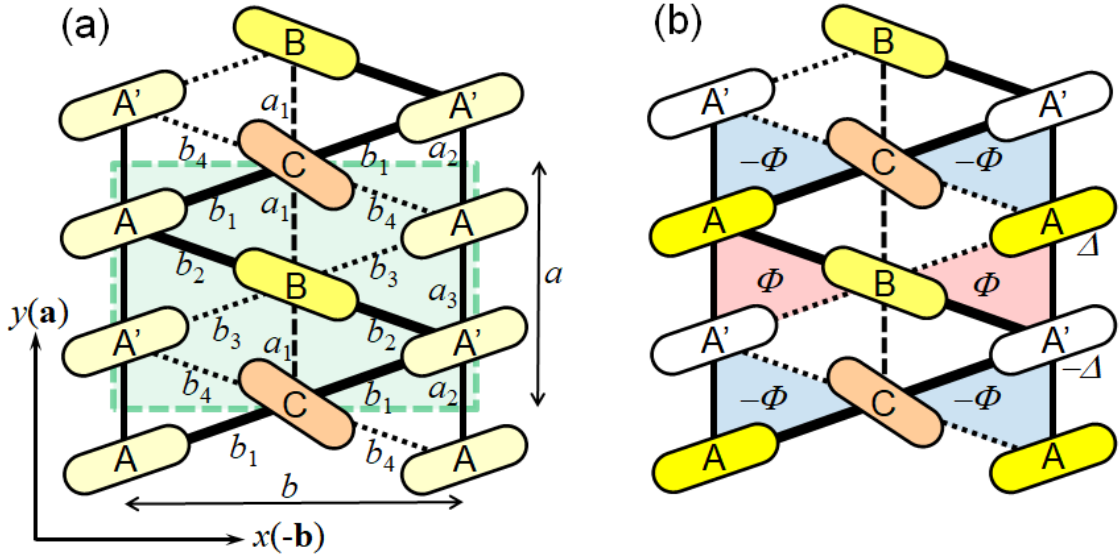

**FIG. 1.** (color online)

(a) Schematic of lattice structure of a conducting layer of $\alpha$-(BEDT-TTF)$_2$I$_3$. There are four BEDT-TTF sites dubbed A, A', B and C in a unit cell indicated by dashed rectangle. The transfer integrals are also indicated. (b) Model of the $\alpha$-(BEDT-TTF)$_2$I$_3$ lattice with potential and magnetic modulations in a unit cell. Site potentials are introduced to sites A and A', and magnetic fluxes are introduced to plaquettes $\triangle$BAA' and $\triangle$CAA'.

**Figure 2** (Osada)

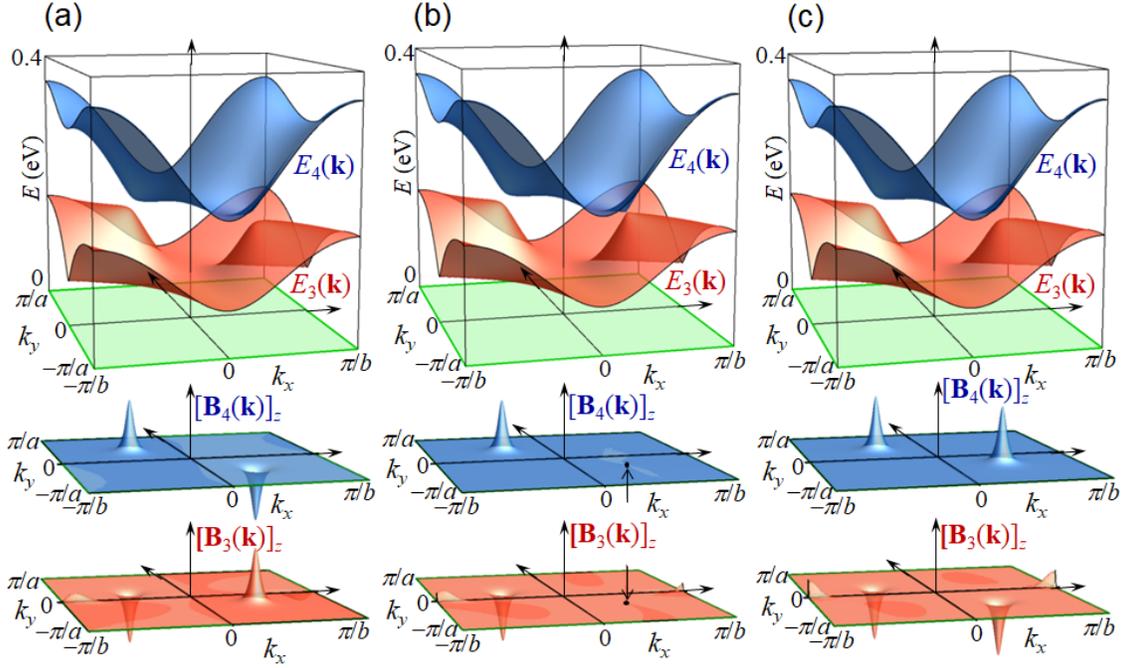

**FIG. 2.** (color online)

Dispersion ($E_3(\mathbf{k})$ and $E_4(\mathbf{k})$) and Berry curvature ($\mathbf{B}_3(\mathbf{k})$ and $\mathbf{B}_4(\mathbf{k})$) of the valence and conduction bands, respectively, for (a) $\Delta = 0.02$ eV, $\Phi = 0$ (ordinary insulator), (b) $\Delta = 0.01$ eV, $\Phi = 0.0128\ h/e$ (2D Dirac semimetal), and (c) $\Delta = 0$, $\Phi = 0.025\ h/e$ (Chern insulator)

**Figure 3** (Osada)

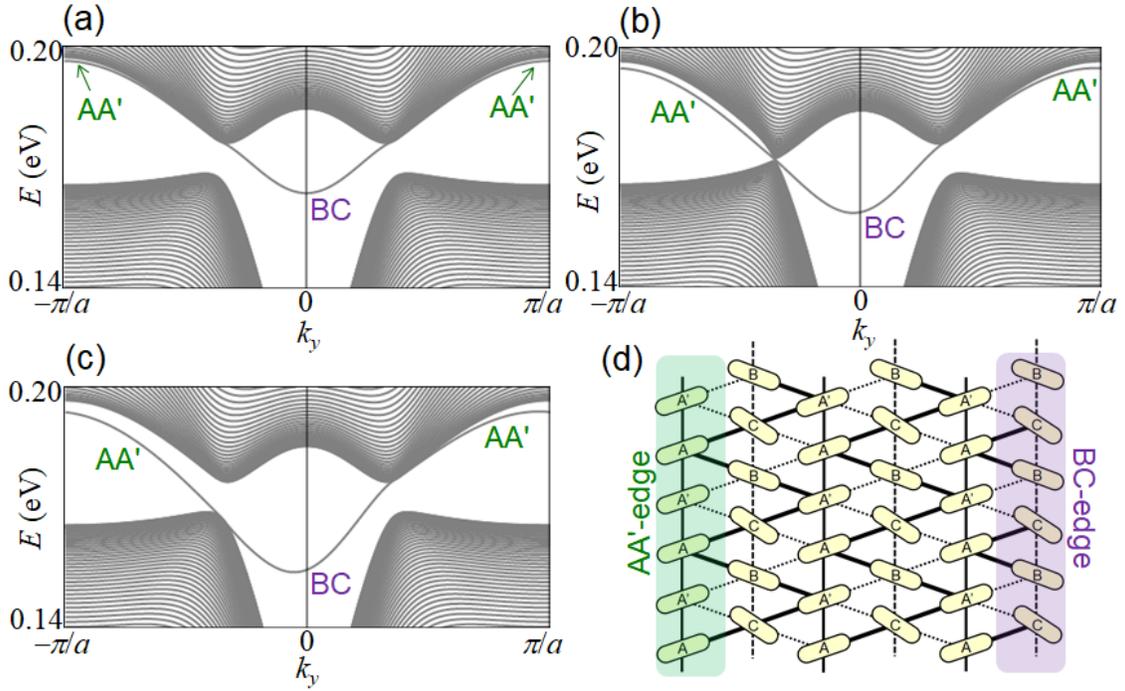

**FIG. 3.** (color online)

Energy dispersion of the nanoribbon of α-(BEDT-TTF)$_2$I$_3$ parallel to the crystalline **a**-axis for (a) $\varDelta = 0.02$ eV, $\varPhi = 0$ (ordinary insulator), (b) $\varDelta = 0.01$ eV, $\varPhi = 0.0128\ h/e$ (2D Dirac semimetal), and (c) $\varDelta = 0$, $\varPhi = 0.025\ h/e$ (Chern insulator). (d) schematic of the nanoribbon of α-(BEDT-TTF)$_2$I$_3$ with the AA'-edge and BC edge. The branches labeled AA' and BC in the bulk gap in (a), (b), and (c) correspond to the edge states along the AA'-edge and BC-edge in (d), respectively.

**Figure 4** (Osada)

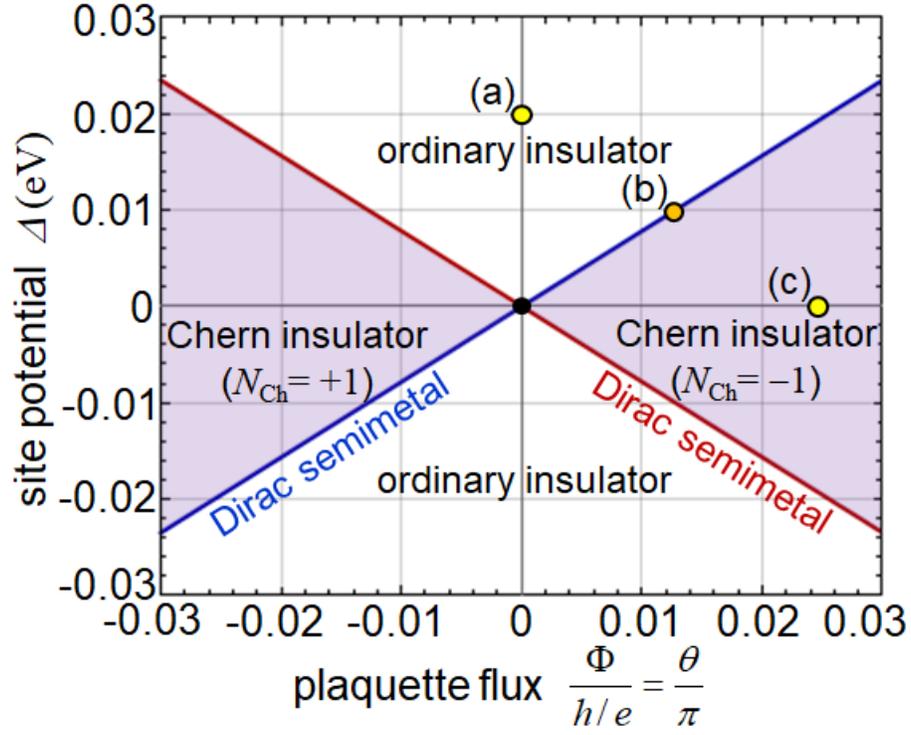

**FIG. 4.** (color online)

Phase diagram of 2D $\alpha$-(BEDT-TTF)$_2$I$_3$ lattice with intracellular potential and magnetic modulations. The points labeled (a), (b), and (c) correspond to parameters appearing in Fig. 2 and Fig. 3.